\documentclass[twocolumn]{aastex631}

\usepackage{xcolor}
\usepackage{amsmath}
\usepackage{hyperref}
\usepackage{xspace}
\usepackage{natbib}
\usepackage{graphicx}
\newcommand{\simba}{{\sc Simba}\xspace}
\graphicspath{{./figures/}{../figures/}}

\newcommand{\lya}{Ly$\alpha$}

\shorttitle{AGN Effects on the \lya~Flux Power Spectrum}
\shortauthors{Tillman et al.}

\begin{document}

\title{The Effects of AGN Feedback on the Lyman-$\alpha$ Forest Flux Power Spectrum}

\author[0000-0002-1185-4111]{Megan Taylor Tillman}
\affiliation{Department of Physics and Astronomy, Rutgers University,  136 Frelinghuysen Rd, Piscataway, NJ 08854, USA}

\author[0000-0001-5817-5944]{Blakesley Burkhart}
\affiliation{Department of Physics and Astronomy, Rutgers University,  136 Frelinghuysen Rd, Piscataway, NJ 08854, USA}
\affiliation{Center for Computational Astrophysics, Flatiron Institute, 162 Fifth Avenue, New York, NY 10010, USA}

\author[0000-0002-8710-9206]{Stephanie Tonnesen}
\affiliation{Center for Computational Astrophysics, Flatiron Institute, 162 Fifth Avenue, New York, NY 10010, USA}

\author[0000-0001-5803-5490]{Simeon Bird}
\affiliation{University of California, Riverside, 92507 CA, U.S.A.}

\author[0000-0003-2630-9228]{Greg L. Bryan}
\affiliation{Center for Computational Astrophysics, Flatiron Institute, 162 Fifth Avenue, New York, NY 10010, USA}
\affiliation{Department of Astronomy, Columbia University, 550 W 120th Street, New York, NY 10027, USA}

\begin{abstract}
We study the effects of AGN feedback on the \lya~forest 1D flux power spectrum (P1D). 
Using the \simba cosmological-hydrodynamic simulations, we examine the impact that adding different AGN feedback modes has on the predicted P1D.
We find that, for \simba, the impact of AGN feedback is most dramatic at lower redshifts ($z<1$) and that AGN jet feedback plays the most significant role in altering the P1D.
The effects of AGN feedback can be seen across a large range of wavenumbers ($1.5\times10^{-3}<k<10^{-1}$ s/km) changing the ionization state of hydrogen in the IGM through heating. AGN feedback can also alter the thermal evolution of the IGM and thermally broaden individual \lya~absorbers.
For the \simba model, these effects become observable at  $z \lesssim 1.0$.
At higher redshifts ($z>2.0$), AGN feedback has a $2$\% effect on the P1D for $k<5\times10^{-2}$ s/km and an $8\%$ effect for $k>5\times10^{-2}$ s/km. We show that the small-scale effect is reduced when normalizing the simulation to the observed mean flux. On large scales, the effect of AGN feedback appears via a change in the IGM temperature and is thus unlikely to bias cosmological parameters. The strong AGN jets in the \simba simulation can reproduce the $z>2$ \lya~forest.
We stress that analyses comparing different AGN feedback models to future higher precision data will be necessary to determine the full extent of this effect. 

\end{abstract}


\section{Introduction}\label{s:Introduction}

Most of the baryons in the Universe reside within the space between galaxies and galaxy clusters known as the intergalactic medium (IGM).
Generally, intergalactic gas is divided into two phases: cool diffuse gas ($T
<10^4$ K) and the warm-hot IGM \citep[WHIM;][]{Dave2001}.
At the present day, about $30\pm10$\% of all gas can be seen in \lya~absorption (HI), which makes up much of the cool diffuse phase \citep[for a comprehensive review of the IGM see][]{McQuinn:2016}.
The WHIM consists of gas that has been shock-heated as a result of nonlinear structure formation and is observable largely in highly ionized tracers such as OVI.
Additionally, heating by outflows from galactic feedback can increase the fraction of gas in the WHIM by tens of percent \citep{Christiansen:2020,Burkhart_2022,Tillman:2023AJL}.
Apart from OVI and HI, other absorption lines can probe structures in or around the circumgalactic medium and their galaxies \citep{Tumlinson:2017}.

The \lya~forest specifically has been a useful tool for both astrophysical and cosmological studies for decades. 
At high redshift ($z\gtrsim 2$, but after HI reionization) the \lya~forest is observable using ground-based telescopes leading to a large archive of observational data to utilize along with ongoing surveys collecting new data \citep[e.g.\ with BOSS, eBOSS, SDSS III and IV, Keck/HIRES, and VLT/UVES;][]{Vogt:1994HIRES, Dekker:2000UVES, Viel:2008, Eisenstein:2011SDSSIII, Dawson:2013BOSS, Viel:2013aKeck/HIRES,  Dawson:2016eBOSS, Pieri:2016WEAVE-SQO, DESI:2016, Blanton:2017SDSSIV, Walther:2018VLT/UVES}.
\lya~absorbing gas tends to reside in under-dense regions at higher redshifts ($z\gtrsim 4$) but by $z=2$ the absorbers arise mostly from gas in mild overdensities  \citep[$\Delta \sim 1 - 10$,][]{Lukic:2015}.
As a good tracer of dark matter, the high-$z$ forest is useful for cosmological studies such as exploring alternative dark matter models \citep{Armengaud:2017, Irsic:2017} and estimating cosmological parameters and the matter power spectrum \citep{Viel+Haehnelt:2006, Chabanier:2019a, Bird:2023, Fernandez:2023}.
At lower redshifts, the IGM temperature is expected to rise from $T_0\sim7500$ K at $z\sim 4.5$ \citep{Boera:2019} to $T_0\sim14750$ K due to HeII reionization around $z\sim 3$ \citep{Gaikwad:2021}, but by $z=0$ the temperature is expected to drop to $T_0\sim 5000$ K \citep{McQuinn:2016, UptonSanderbeck:2016}.
As such, the \lya~forest is an ideal probe for studying how the temperature of the cool diffuse IGM evolves from reionization to the present day.

Recent works have focused on the potential for galactic feedback to heat the low-$z$ \lya~forest \citep{Viel:2017, Gurvich2017, Tonnesen:2017}.
Specifically, active galactic nuclei (AGN) feedback, which is expected to be more prominent at lower redshifts, could produce heating on Mpc scales and thus potentially affect the \lya~forest \citep{BroderickI:2012, ChangII:2012}.
Currently, feedback models are generally calibrated to the observed galaxy stellar mass function, and multiple AGN models have produced physically reasonable galaxies in this regard.
Therefore, it will be necessary to seek additional independent observational constraints to distinguish between AGN feedback models and the \lya~forest may provide such a novel constraint.
To develop the \lya~forest as such a constraint, simulations with identical setups but different AGN feedback models will be required. 
The CAMELS simulations \citep{CAMELS-public} may provide the best data set for such an analysis; however, ideally the simulations would use the same hydrodynamic solvers too.
Furthermore, obtaining a better understanding of the potential impact that AGN feedback can have on \lya~forest statistics will be necessary to perform a detailed analysis.
In this work we will focus on the latter, following up on recent works with similar goals.

Many recent studies on the low-z \lya~forest have explored the impact of AGN feedback on the IGM, but there is not yet a concrete method for observationally disentangling the effects of AGN from other heating and ionization sources \citep{Nasir:2017, Christiansen:2020, Chabanier:2020, Burkhart_2022, Khaire:2022, Tillman:2023AJL, Tillman:2023AJ, Dong:2023, Khaire:2023, Dong:2024, Khaire:2024}.
In many of these works, the effects of AGN feedback on the column density distribution function was found to be highly degenerate with re-scaling the assumed UVB.
Additionally, no simulation matches the observed $b$-value distribution at $z<$1. 
Out of the \lya~forest summary statistics that are typically analyzed, \citet{Burkhart_2022} and \citet{Khaire:2024} found that the 1D transmitted flux power spectrum (P1D) may provide the best opportunity to study the unique effects of feedback.

On small physical scales, the P1D can give us insights into the thermal state of the IGM, which is useful for studying reionization \citep{Zaldarriaga:2001, Meiksin:2009, Lee:2015, McQuinn:2016, Boera:2019, Zhu:2021} and the thermal history of the IGM \citep{Viel+Haehnelt:2006, Bolton:2008, Becker:2011, Rudie:2012, Hu:2023}.
Because the \lya~forest traces small scale structures in the Universe, the P1D can also be used to make constraints on cosmological parameters such as the mass of neutrinos \citep{Palanque-Delabrouille:2015a, Palanque-Delabrouille:2015b, Yeche_2017, Palanque-Delabrouille:2020}, the nature of dark matter \citep{Yeche_2017,Palanque-Delabrouille:2020,Villasenor:2023, Baur:2016}, and more \citep[e.g.\ dark radiation, sterile neutrinos;][]{Baur:2017, Rossi:2017, Garny:2018, Rossi:2020}. 
Whether these studies are conducted through the study of the P1D alone or through cross-correlation with other statistics, understanding how to marginalize out the effects of AGN feedback is necessary.

The role of AGN feedback in setting the P1D, as seen through hydrodynamical simulations, has the potential to bias the measured cosmological values \citep{Viel:2013,Chabanier:2020,Burkhart_2022}.
Better understanding the potential effects of AGN on the P1D will be vital in the era of precision cosmology and has even been discussed as a partial solution to the $S_8$ tension \citep{Chen:2022, Schneider:2022, Arico:2023, Elbers:2024, Terasawa:2024}.
This tension arises from disagreements between measurements of $S_8 = (\Omega_{m}/0.3)^{0.5} \sigma_8$ derived from the Cosmic Microwave Background versus measurements made from other probes of inhomogeneity such as the \lya~forest and weak lensing \citep[via the DES or KiDS surveys;][]{DES:2005,KiDS:2013,Davies:2021, Amon:2022, Fernandez:2024}.

An important challenge for the use of the \lya~forest to address scientific questions is the availability of observational data, specifically at $z \lesssim2$. After this time, the \lya~transition wavelength is shifted into the far-ultraviolet (FUV) band which requires the use of space-based spectrographs to observe.  Additionally, that spectrograph needs to have high enough resolution to resolve the individual sightlines that constitute the \lya~forest. Some instruments currently exist that can collect this data such as Hubble Space Telescope's (HST) Cosmic Origins Spectrograph (COS) \citep{Danforth:2005}. However, HST can not observe the \lya~forest at redshifts higher than $z\sim0.5$, leaving a large gap in the available observational data on the \lya~forest (from $z\sim 0.5$ to $z\sim1.8$) that we dub the ``\lya~forest blind spot''. While much can still be accomplished with currently available catalogs of \lya~forest data \citep[such as][]{Danforth:2016, Khaire:2019}, the collection of additional data in the observational blind spot will be necessary to further improve our understanding of the IGM. 

In this work, we analyze the \simba simulations to explore the potential effects of AGN feedback on the \lya~forest P1D across a wide redshift range ($0.03<z<3$). Predictions for observations in the blind spot will be necessary to motivate future missions, and therefore we analyze several redshift bins in that range despite the lack of current data.
We utilize four \simba simulations that are all identical apart from the AGN feedback model that is implemented.
Comparing the predicted P1D from each simulation can reveal the effects of AGN feedback on the \lya~forest. 
We then compare the relative changes seen in the P1D to the precision of currently available observational data to determine if these effects are observable or could impact cosmological predictions.

The outline of this paper is as follows: In \textsection \ref{s:Methods} we discuss the simulations we analyze, along with the statistics we explore and how they are generated.
In \textsection \ref{s:Results} we analyze the results from our analysis across three different redshift bins.
In \textsection \ref{s:Discussion} we discuss the results in a broader context and consider caveats resulting from the use of simulated data and observational limitations.
Finally in \textsection \ref{s:Conclusion} we summarize the results of this work.

\section{Methodology}\label{s:Methods}
    
    \subsection{Simulations}\label{ss:simulations}

        In this section we present a brief review of the main aspects of the code for the simulations analyzed herein, with an emphasis on how the AGN feedback model is implemented. 
        For more details on how the \simba simulations are constructed, refer to \citet{dave:2019}.
        \simba is the next generation of the {\sc Mufasa}\xspace cosmological galaxy formation simulations \citep{dave:2016}. These simulations are run with GIZMO’s meshless finite mass hydrodynamics \citep{hopkins:2015}, and employ state-of-the-art subgrid physical processes to produce realistic galaxies. 
        GIZMO bases its gravity solver on GADGET-3 \citep{Springel:2005} and evolves dark matter and gas together including pressure forces and gravity. 
        The shocks in GIZMO are handled via a Riemann solver without artificial viscosity.
        The following cosmological parameters are used: $\Omega_m = 0.3$, $\Omega_\Lambda$ = 0.7, $\Omega_b = 0.048$, $H_0 = 68$ km/s/Mpc, $\sigma_8 = 0.82$, and $n_s = 0.97$.
        Photo-ionization, photo-heating, and radiative cooling are implemented using GRACKLE-3.1 \citep{Grackle:2017} assuming ionization equilibrium but not thermal equilibrium.  
        GRACKLE is used assuming the \citet{Haardt:2012UVB} ionizing background, modified to account for self-shielding \citep{Rahmati:2013}.

        In the \simba simulations, AGN feedback is modeled in two main modes with a third mode that is reliant on one of the other two. There is a radiative mode at high Eddington ratios ($\eta = \dot{M}_{BH}/\dot{M}_\textnormal{Edd}$ where $\dot{M}_\textnormal{Edd} = L_\textnormal{Edd}/\epsilon_r c^2$), a jet mode at low Eddington ratios, and an X-ray mode that accompanies jets with maximum outflow velocities.
        The radiative and jet mode are implemented as kinetic outflows \citep{DAA:2017a}, while the X-ray mode is modeled as X-ray photon energy feedback \citep[via thermal energy injection as in][]{Choi:2012}. 
        
        When the SMBH is accreting mass at high Eddington ratios, the radiative mode drives multi-phase gas in winds at velocities of $\sim 500-1500$ km/s. 
        This mode typically dominates at higher redshifts and in lower-mass galaxies.
        The jet mode feedback occurs for SMBHs with Eddington ratios of $\eta$ $< 0.2$ and masses of $M_{\rm BH} \geq 10^{7.5} M_\odot$.
        The jet velocity is calculated as the radiative mode velocity with an additional boost that gets larger with lower accretion rates.
        The largest boost a jet can have is 7000 km/s and occurs when $\eta\leq0.02$.
        Overall, this model can produce jets that reach maximum velocities around $8000$ km/s.

        The jets eject gas that is heated to the virial temperature of the host galaxy. 
        The ejected gas is decoupled from hydrodynamics and cooling for a (short) period of time that scales with the Hubble time, meaning that jets can travel up to $\sim$ 10 kpc until depositing their thermal energy in the surrounding medium.
        The feedback is expelled as highly collimated bipolar jets that align with the angular momentum vector of the galactic disk.

        When AGN jets have the maximum velocity boost, and the galaxy has a gas to stellar ratio of $M_\textnormal{gas}/M_* < 0.2$, X-ray feedback also affects the surrounding medium.
        Non-ISM gas is heated by the X-ray feedback with a rate decreasing with distance from the SMBH.
        Star-forming gas is heated as well, but 50\% of the energy injected is applied as a radial kick outwards instead. 
        Additional information on the \simba simulation sub-grid models can be found in \citet{dave:2019}. 

        The simulations we analyze here are a subset of the \simba simulations that vary which feedback modes are implemented \citep[explored in a similar context in][]{Christiansen:2020}. 
        These simulations have a box length of 50 Mpc/$h$ and $2\times512^3$ particles.
        There is a \simba simulation with no feedback at all, but this work uses the stellar-feedback-only (i.e., in the form of supernova winds) simulation as the baseline since AGN feedback is the primary focus.
        We refer to this simulation in text and plots as \textit{SW Only}.
        Stellar feedback (in \simba) is expected to have minimal impact on IGM statistics as a whole.
        The role of stellar winds in cycling gas out of the host galaxy will have some impact, especially closer to halos and in the circumgalactic medium of the host; however, it is unlikely that the stellar feedback will have the ability to produce effects far into the IGM.
        
        We also analyze simulations that introduce the AGN feedback modes one by one. 
        The first mode introduced is the radiative feedback mode, so the \textit{+Rad} simulation has stellar feedback and radiative AGN feedback.
        The \textit{+Rad+Jet} simulation has stellar feedback, AGN radiative feedback, and AGN jet feedback.
        Finally the \textit{+Rad+Jet+X} simulation has all forms of feedback and is essentially the fiducial \simba simulation just run in a smaller box.
        Given previous work, of the AGN feedback modes, it is expected that the radiative mode and the X-ray mode will have a lower impact on the IGM than that of the jet mode \citep{Christiansen:2020, Tillman:2023AJ}.
        The jet feedback effects are expected to be far reaching, with the ability to heat gas tens of kpc away from the galaxy and effects that can escape the host halo.
        The radiative and X-ray modes are both local feedback effects and are expected to have the biggest impact on star formation and SMBH fueling.
        Apart from the differences in the feedback models, the four simulations are identical with the same box size, particle resolution, initial conditions, and cosmology.

        This makes these four simulations an ideal testbed to analyze the effects these individual AGN feedback modes have on the IGM.
        \textbf{However, we stress that the comparison of these four simulations does not represent a comparison between four different physically reasonable AGN feedback models, as each simulation produces different galaxy statistics such as in their predicted stellar mass function (SMF) and star formation rates.
        However, comparisons between a model with and without X-ray feedback (\textit{+Rad+Jet} vs.\ \textit{+Rad+Jet+X}) produce similar SMFs, at least within observational error bars, and could represent a case where effects on the \lya~forest could help constrain AGN feedback \citep{dave:2019}.}

        \subsection{Generating Synthetic Spectra}\label{ss:Spectra}

        We generate synthetic spectra for the \simba simulations analyzed here using the publicly available \textit{fake\_spectra}\footnote{\url{https://github.com/sbird/fake_spectra}}\footnote{Frozen version used herein at \url{https://github.com/megantillman/fake_spectra_mod/tree/master}} code outlined in \citet{Bird:2015, Bird:2017} with MPI support from \citet{Qezlou:2022}. 
        From simulation snapshots the code generates and analyzes mock spectra. The \textit{fake\_spectra} package is fast, parallel, and is written in C++ and Python 3 with the user interface being Python-based. 

        We generate 5,000 sightlines randomly placed in each simulation box; a number found to be sufficient for avoiding variations due to sampling \citep{Tillman:2023AJ}. 
        Noise is not added to the spectra generated from the simulation box since we want to understand the data in the limit of high signal to noise ratio.
        We generate the optical depth $\tau$ from which the flux is obtained via $F(v) = \exp{(-\tau)}$. 
        Optical depths larger than $\tau = 10^{6}$ are considered to be DLA regions and are filtered out of the spectra before computing the P1D.

\subsection[Analysis: The 1D Lyman-alpha Flux Power Spectrum]{Analysis: The 1D \lya~Flux Power Spectrum}\label{ss:P1D}

        \begin{figure*}
            \centering
            \includegraphics[width = \linewidth, trim=0.0cm 0.0cm 0.0cm 0.0cm, clip=true]{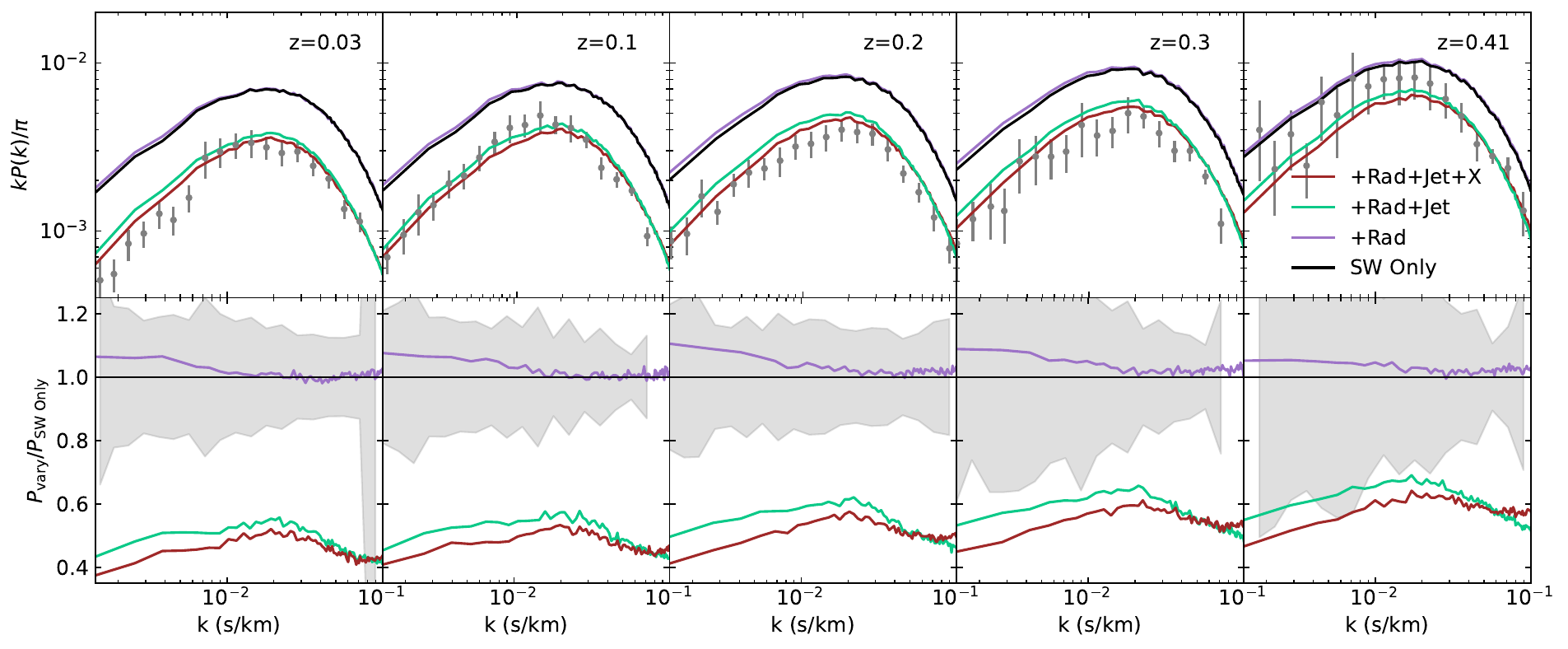}
            \caption{\textbf{Top row:} \lya~forest P1D for the \simba simulations at low-$z$. Grey points with errorbars are observational data from \citet{Khaire:2019}. \textbf{Bottom row:} P1D ratios of the plots above. For each simulation, the ratio is taken of the simulation variant to the stellar feedback only (\textit{SW Only}) simulation. The shaded region represents the error on observational data from \citet{Khaire:2019}, assuming the \textit{SW Only} simulation was ground truth.}
            \label{fig:PS_lowz}
        \end{figure*}

        From the flux obtained from our simulated sightlines we calculate the one-dimensional \lya~flux power spectrum. The \lya~P1D describes the variance of the \lya~flux along the line of sight across different physical scales. The P1D is defined as the average (over spectra) amplitude of the 1D Fourier transform of the flux fluctuations $\delta_F(v)$, where

        \begin{equation}
            \delta_F(v) \equiv \frac{F(v)-\langle F\rangle}{\langle F\rangle}.
        \end{equation}

        \begin{figure*}[t]
            \centering
            \includegraphics[width = \linewidth, trim=0.0cm 0.0cm 0.0cm 0.0cm, clip=true]{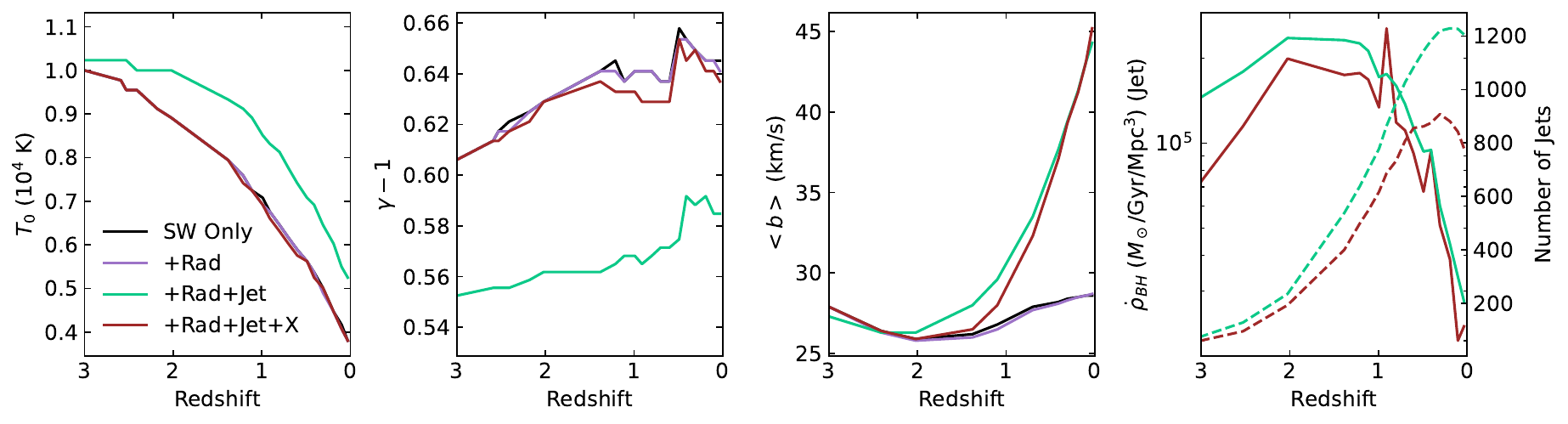}
            \caption{From left to right the plots in this Figure are \textbf{(1)} The evolution of the temperature at the cosmic mean density ($T_0$) of the \simba simulations. \textbf{(2)} The evolution of $\gamma -1$. \textbf{(3)} The evolution of the average $b$-value of \lya~absorbers. \textbf{(4)} The evolution of the accretion rate density of jet mode SMBHs (solid lines) and the total number of jets (dashed lines).}
            \label{fig:thermalhistory}
        \end{figure*}

        \noindent Above, $\langle F \rangle$ is the mean transmitted flux over the sightline (averaged over all sightlines at a given redshift) and $v$ is the position in the \lya~forest sightline in units of km/s. The corresponding P1D is then in terms of wavenumber $k$ (units of s/km). 
        For the P1D generated from simulations, we analyze results down to a $k$ value corresponding to the physical length of the box at each given redshift. 
        For a 50 Mpc/h box, like the \simba simulations analyzed herein, the minimum wavenumber is on the order of $k\approx10^{-3}$ s/km. 
        Fluctuations due to cosmic variance are to be expected due to the limited box size,particularly at the largest scales \citep{Pedersen:2023, Bird:2023}.
        Fluctuations due to box size are expected to be on the order of 2\% and up to 5\% for the smallest $k$ values.
        Results are analyzed up to a maximum $k$ value of 0.1 s/km.
        Particle resolution convergence must also be considered. 
        The simulations analyzed herein have a similar particle resolution as TNG100-2 which was found to be converged within $\sim 8$\% of the higher resolution TNG100-1 at $z=0.1$ \citep{Burkhart_2022}.
        Convergence within 10\% has been found for similar box size and resolution simulations of the \lya~forest that do not contain feedback \citep{Arinyoiprats:2015}, and this is expected to become worse at higher redshifts when lower over-densities dominate the forest.
        As such we continue our analysis under the assumption that our results are converged for our adopted box size and resolution to within 10\%.
        The reasoning for this conclusion and convergence in general are further discussed in appendix \ref{ss:convergence}.

    \subsection[Analysis: Lyman-alpha Forest and the IGM Thermal History]{Analysis: \lya~Forest and the IGM Thermal History}

        Post-reionization, the HI in the IGM that constitutes the \lya~forest is optically thin and is in photoionization equilibrium \citep[given HI and HeII reionization are not in progress, ][]{Gaikwad:2019, Kusmic:2022}. 
        This allows for assumptions that greatly simplify the study of the IGM through the \lya~forest.
        For example, the fluctuating Gunn-Peterson approximation (GPA) describes how the optical depth of neutral hydrogen in the \lya~forest scales with photoionization rate, temperature, and density \citep[$\tau \propto \Gamma_{HI}^{-1}T^{-0.7} n_{H}^2$][]{Gunn:1965,Croft:1998}.
        This relation shows the relative simplicity of the physics that governs the \lya~forest and suggests which mechanisms are important to focus on.

        The photoionization rate scaling makes the \lya~forest a good probe for the ionizing background \citep{Haardt:2012UVB, Gaikwad:2017UVB, Khaire:2019UVB, Puchwein:2019UVB, Faucher-Giguere:2020UVB}.
        Values for photoionization and photoheating rates can be predicted from the absorption properties of HI and HeII in the IGM, and in turn these values help set the ionization fraction and temperature of the IGM in analytic models and simulations.
        Particularly at high-$z$ ($z>2$), it is predicted that the balance between photoheating and adiabatic cooling sets the temperature of the \lya~forest.
        That balance leads to a power law relation seen in the temperature-density phase diagram describing the temperature of the diffuse gas in the IGM as $T = T_0 \Delta ^{\gamma -1}$ \citep[where $T_0$ is the temperature at the mean cosmic gas density, $\Delta$ is the overdensity, and $\gamma - 1$ is the slope,][]{Hui+Gnedin:1997,McQuinn+Sanderbeck:2015}.
        There exists scatter in this relation due to heating from other mechanisms and this relation does not apply to WHIM gas. 
        In fact, the state of the WHIM does not appear to affect the \lya~forest in any significant way \citep{Hu:2024}.
        The evolution of $T_0$ and $\gamma-1$ are used to study the evolution of the thermal history of the IGM and will be used as such herein.
        While these values do not fully describe the \lya~forest on their own, they reveal information about the environment in which the \lya~forest gas resides.
        For example, pressure smooths the structure of the forest, and differences in smoothing between simulations can be studied through the evolution of $T_0$.

    \subsection[Analysis: The Doppler Parameter of Lyman-alpha Absorbers]{Analysis: The Doppler Parameter of \lya~Absorbers}\label{ss:bparam}

        The shape of the P1D is affected by multiple mechanisms including the temperature and turbulence of individually fit \lya~absorption lines which can be described by their Doppler parameter or $b$-value.
        For the case of pure thermal broadening in absorbers, the $b$-value scales as

        \begin{equation}
            b\propto \sqrt{\frac{2kT}{m_H}}
        \end{equation}

        \noindent where $k$ is the Boltzman constant, $m_H$ is the mass of hydrogen, and $T$ is the temperature of the absorber.
        An absorber's $b$-value will also scale with relative velocities, such as the root mean squared turbulent velocity of the absorbing gas, as well as the peculiar and Hubble velocity gradients across the absorber.
        An increase in the average $b$-values will result in a suppression of the high-$k$ end of the P1D (on the scale of the $b$-value in $k$ space).
        Since the effects of thermal broadening on the P1D are degenerate with the effects of pressure smoothing \citep{Nasir:2016}, it will be enlightening to look at the average $b$-value and $T_0$ simultaneously when analyzing the results herein.
        Increases in $b$ that do not correspond with increases in $T_0$ suggests that absorber broadening is due to turbulence driven by feedback rather than heating.
        It may also suggest that those absorbers still lie in the cool diffuse regime of the temperature density phase space but not on the power law relation described by $T_0$.

        The $b$-value of absorbers in our synthetic spectra are extracted through an automated Voigt fitting algorithm included in the \textit{fake\_spectra} package. 
        The algorithm is modified slightly for our use to more easily fit smaller peaks in the spectra \citep[see][]{Tillman:2023AJ}. 
        We do not convolve the spectra with a point spread function since we do not intend to compare the derived $b$-values to observations.
        In this work, we focus on the relative changes of the median $b$-value between the simulations with varying AGN feedback.
        Analysis of how both $b$ and  $T_0$ evolve over time gives a clearer picture than either alone of how the thermal and turbulent state of the \lya~forest is affected by AGN.

\section{Results}\label{s:Results}

    In this work three main redshift regimes are analyzed: (1) a low redshift $(z<0.5)$ regime where observations from current spaced-based telescopes exist, (2) an intermediate redshift $(0.5<z<2)$ regime that lacks observational data, and (3) a high redshift $(z>2)$ regime where ground-based telescopes collect \lya~forest data largely used for cosmological studies.
    Understanding what effects AGN feedback has on the \lya~forest at these different epochs will reveal where higher precision data is needed both to constrain future feedback models and to marginalize out feedback effects.

    \begin{figure*}
        \centering
        \includegraphics[width = 0.8\linewidth, trim=0.0cm 0.0cm 0.0cm 0.0cm, clip=true]{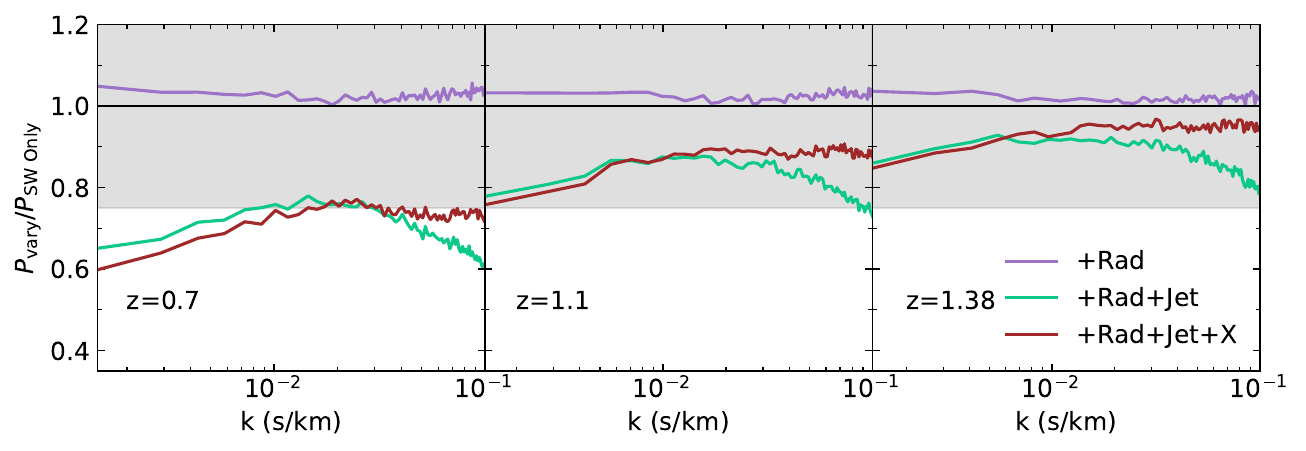}
        \caption{P1D ratios for intermediate redshifts occupying the \lya~forest blind spot. The effects of AGN feedback in \simba becomes observable to 25\% accuracy (the shaded grey region) below $z\approx 1$.}
        \label{fig:PSratio_interz}
    \end{figure*}

    The P1D probes different physical scales and thus different aspects of the \lya~forest.
    The high-$k$ end of the P1D is a probe for the thermal state of the \lya~forest and will be more sensitive to changes in temperature (e.g.\ $T_0$).
    The low-$k$ end, as a large scale probe, will be more affected by the amount of HI in the \lya~forest and changes to large scale structures.

    \subsection{Low Redshift Effects}\label{ss:lowz}
        At low-$z$ ($z<0.5$) the IGM is less sensitive to the details of H and He reionization, which completed long ago.
        As the IGM settles into ionization equilibrium and cools, we can more confidently analyze the high-$k$ end of the P1D as a probe for the instantaneous thermal state of the IGM.
        While this thermal state will still depend on the photoheating and photoionizing rates of the UVB, we should expect galactic feedback effects to be more clear at these redshifts due to the abundance of SMBHs.
        Figure \ref{fig:PS_lowz} shows the P1D and P1D ratios of the \simba simulations analyzed, for several redshift bins at $z<0.5$ compared to observational data from \citet{Khaire:2019} derived from the low-z IGM HST/COS survey by \citet{Danforth:2016}.

        The P1D ratios are the ratios of the different \simba simulations to that of the \simba simulation containing only stellar feedback (\textit{SW Only}). 
        This was chosen to analyze what happens when feedback modes are added in one by one.
        When adding in radiative AGN feedback (\textit{+Rad}) there is an increase in power on the low-$k$ end of the P1D. 
        The increase seen is most likely associated with less heating from stellar feedback. 
        Indeed, the radiative feedback mode is implemented to suppress star formation and there is a reduction in SFRs with the introduction of AGN radiative winds \citep{Scharre+2024}. 
        This results in less energy released in the form of stellar feedback overall.
        The change to the P1D is nevertheless less than 10\% (not observable with current data), and these changes are only seen at the lowest wavenumbers ($k\lesssim 10^{-2}$ s/km).
        Comparable changes at higher wave numbers would be due to thermal broadening of \lya~absorbers or a change in the thermal history of the IGM.
        We can check if either of these aspects changes by looking at the average $b$-value of the predicted \lya~forest and looking at $T_0$ and $\gamma -1$.
        
        The first three panels in Figure \ref{fig:thermalhistory} show the evolution of $T_0$, $\gamma-1$, and the average $b$-value of each of the simulations.
        There is no change to these values when adding in AGN radiative feedback.
        This implies that the impact of stellar feedback that causes the change seen at lower $k$ is due to ionization through heating of \lya~absorbers primarily in the circumgalactic medium.
        Feedback effects on these absorbers will still impact \lya~statistics but are unlikely to appear in IGM specific values such as $T_0$ and $\gamma$.
        We refrain from comparing $T_0$ and $\gamma$ with observational data, since \simba uses an older UVB model from \citet{Haardt:2012UVB}.
        In this model, HeII reionization occurs earlier than current observations imply, so the results in the two left panels of Figure \ref{fig:thermalhistory} are not expected to match observations and should only be used to compare between the different \simba simulations.

        \begin{figure*}
            \centering
            \includegraphics[width = 0.8\linewidth, trim=0.0cm 0.0cm 0.0cm 0.0cm, clip=true]{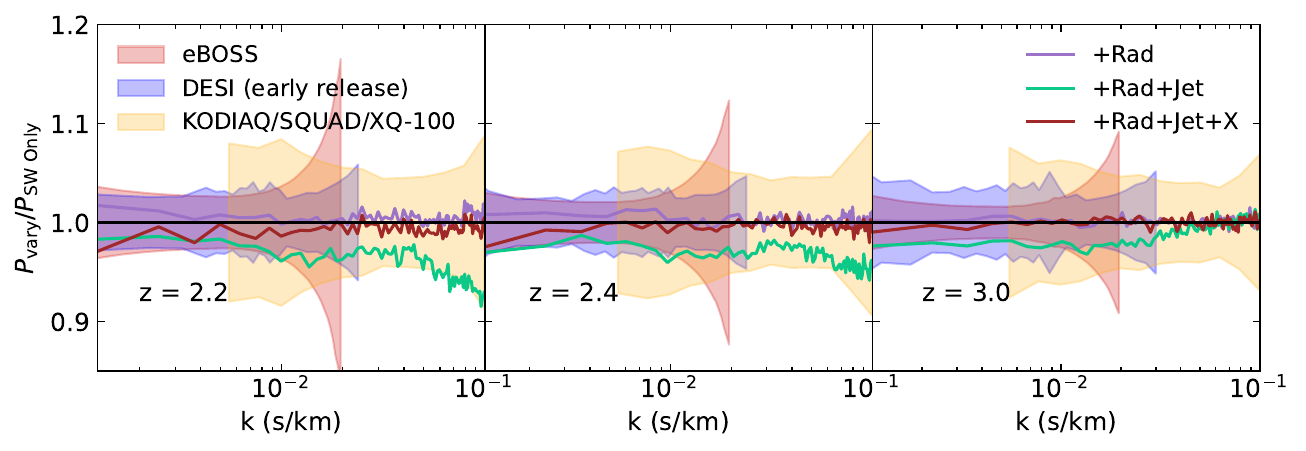}
            \caption{P1D ratios for higher redshifts with shaded regions representing current observational precision. The red shaded region represents the error from the eBOSS derived P1D \citep{Chabanier+2019}. The blue shaded region represents the error from the DESI (early release) derived P1D \citep{Ravoux+2023, Kara+2024}. 
            The orange shaded region represents the error from P1D derived from a combination of KODIAQ, SQUAD, and XQ-100 surveys \citep{Karacayli:2022}.
            Changes to the P1D due to the AGN feedback model in the \simba simulation are just within the observational error for currently available data.
            However precision is expected to increase with the full DESI survey.}
            \label{fig:PSratio_highz}
        \end{figure*}

        The largest change in the P1D is due to jet feedback (\textit{+Rad+Jet}), dropping the power at all $k$ by 60\%.
        This aligns with results seen in previous works analyzing the effect of AGN feedback in \simba \citep[][]{Christiansen:2020, Sorini:2022, Tillman:2023AJL, Tillman:2023AJ, Scharre+2024}.
        This highlights that aggressive AGN heating can ionize much of the remaining HI in the \lya~forest. 
        The AGN produce shock-heated bubbles that completely erase small-scale structures within the bubble while leaving those outside untouched \citep[see Figure 1 of][]{Tillman:2023AJ}.
        Due to the erasure of small-scale structure from the large-scale impact of the jets, the effect on the P1D manifests as larger changes at low-$k$ and then decrease as $k$ approaches the filtering scale.
        
        On scales smaller than the filtering scale, changes to the P1D are expected to be due to changes in the thermal state of the \lya~forest.
        Indeed, we see an increase in both $T_0$ and the average $b$-value when adding in jets, which correspond to the suppression of the high-$k$ P1D.
        Seeing changes on both large and small physical scales implies not just the erasure of structure through AGN heating, but also the smoothing of remaining structures.

        Temperature-density phase space diagrams show that the jet heating in \simba primarily moves gas from the diffuse IGM phase to the WHIM \citep[Figure 3 in][]{Christiansen:2020}.
        Adding jets increased $T_0$ and decreased the power law index ($\gamma -1$), as seen in Figure \ref{fig:thermalhistory}.
        The jets are unable to do this once X-ray feedback is implemented, implying that X-ray feedback provides some form of self-regulation for the SMBH.
        This can be confirmed by looking at the total accretion rate density of SMBHs in the jet mode (which correlates with the energy released in the form of jets) and the total number of jets in the box.
        The rightmost panel of Figure \ref{fig:thermalhistory} shows that both of the values of interest decrease when adding in X-ray feedback.
        Not only are less SMBHs able to reach the mass threshold required for jet feedback resulting in less jets overall, but the jets that remain release less energy in the form of feedback as well.
        SMBH growth is overall suppressed by the inclusion of the X-ray mode.
        When adding in AGN X-ray feedback (\textit{+Rad+Jet+X}) there is a slight decrease in power at low-$z$. 
        The evolution of $T_0$ largely returns to what it was before adding in jets but there is still an increase in the average $b$-value from the no feedback case. 
        This highlights that the \lya~forest is not necessarily a perfect indicator for the thermal state of all diffuse cool gas in the IGM and vice versa.
        Indeed the \lya~forest is sensitive to a range of temperatures ($10^3\lesssim T \lesssim 10^5$K) and over-densities \citep[$\log \Delta = 0 \pm 2$][]{Hu:2024}.
        Additionally at lower redshifts, \lya~absorbers can have overdensities on the order of $\Delta\sim10$ and even greater which exceeds the range of the $T$-$\rho$ phase space in which the characteristic power law used to describe the thermal state of the IGM resides.
        The X-ray feedback acts to quench star formation in massive galaxies when jet feedback is active, and while this generally occurs more at lower redshifts, it can be seen to quench galaxies up to $z=2$ \citep{Scharre+2024}. 
        
        Since SFRs tend to be lower when including X-rays, the additional loss of power in the P1D ($\sim 10$\%) must be due to AGN heating.
        The X-ray feedback mode acts to self-regulate the growth of its SMBH resulting in less SMBHs producing jets and less growth for SMBHs.
        Therefore there are a few likely explanations for the loss of power: (1) lower accretion rates correspond to faster jets and faster jets can heat further into the IGM, or (2) the X-ray feedback itself sufficiently heats the surrounding medium enough to affect the predicted IGM.
        However, confirmation would require more in-depth analyses on the amount of energy injected into the IGM through AGN feedback and to what scales that energy reaches, which is beyond the scope of this work.
        The important results for our purposes is the following.
        In the \textit{+Rad+Jet} simulation, differences in both pressure smoothing and thermal broadening play a role in suppressing the high-$k$ P1D.
        Whereas in the \textit{+Rad+Jet+X} simulation, the suppression is due primarily to thermal broadening.
    
    \subsection{Intermediate Redshift Effects}\label{ss:interz}
        Figure \ref{fig:PSratio_interz} shows the same P1D ratios at some intermediate redshifts that occupy the redshift range for which observational data is lacking.
        Here, our goal is to forecast the redshift at which changes in the P1D, due to AGN feedback, become observable.
        For reference, we include a shaded region representing a 25\% change as this is the average observational error for currently available data at redshifts $z\lesssim0.4$ \citep[as seen in][]{Khaire:2019}.
        Errors are lower at higher redshifts ($z\gtrsim$2), but at intermediate redshifts a space based instrument is still required so the errors seen in \citet{Khaire:2019} are more relevant.
        Realistically, the error will be different depending on many factors, but we use this point of reference for simplicity and as a prediction. 
        Figure \ref{fig:PSratio_interz} shows the effects of AGN feedback becomes observable at some point below $z\sim 1.0$. 
        This is consistent with results from previous work, where changes in the column density distribution due to, at least in part, AGN feedback become apparent at redshift slightly below $z=1.0$ \citep{Tillman:2023AJ}.
        This highlights the importance of higher sensitivity UV space instrumentation (e.g., Habitable Worlds Observatory) to probe the \lya~forest in the intermediate redshift range of $z\sim 0.5$ to $z\sim 1.8$.
    
        In the same plot, the differences in the P1D for the \textit{+Rad} simulation are smaller than those seen at lower redshifts.
        However, there is still a gain in power on all scales, implying the effects of AGN and stellar feedback are present in the IGM.
        For the \textit{+Rad+Jet} simulation we see a similar behavior as at lower redshifts, but the difference is lower as fewer AGN are producing jets at those times.
        This results in a flattening of the low-$k$ end of the P1D, as we move from low to high $z$, since the impact of jets is lower at earlier times.
        The high-$k$ end maintains additional suppression, consistent with our finding that $T_0$ and the $b$-values remain larger in this simulation.
        However from  Figure~\ref{fig:thermalhistory} we see that the average $b$-value approaches the \textit{SW Only} case at $z\sim2$ while $T_0$ remains high until $z\sim 3$. 
        This implies that pressure smoothing may be the dominant player in the changes we see at those times for the \textit{+Rad+Jet} simulation.
    
        The \textit{+Rad+Jet+X} simulation behaves similarly to the \textit{+Rad+Jet} simulations at these redshifts.
        However, since the $T_0$ evolution of this simulation matches that of the \textit{SW Only} simulation, pressure smoothing does not contribute to suppression at high-$k$. 
        The main mechanism suppressing the high-$k$ end for the \textit{+Rad+Jet+X} simulation is thermal broadening.
        The average $b$-value converges to what is seen in the \textit{SW Only} case at $z\sim1.5$ resulting in a flattening of the high-$k$ P1D.

    \subsection{High Redshift Effects}\label{ss:highz}
        Figure \ref{fig:PSratio_highz} shows the same P1D ratios but at higher $z$. 
        We compare to \lya~P1D observational data derived from \citep[eBOSS,][]{Chabanier+2019,Dawson:2016eBOSS}, a combination of KODIAQ, SQUAD and XQ-100 \citep{XQ-100:2016,KODIAQI:2015,KODIAQII:2017,SQUAD:2019,Karacayli:2022}, and the DESI early release data \citep{Ravoux+2023,Kara+2024,DESIa:2024,DESIb:2024}.
        Current observational data puts our sensitivity of the P1D around 2\% at $z=2.4-3.8$ for $10^{-3}<k<10^{-2}$ s/km with decreasing sensitivity for higher $k$ where up to $k=0.1$ s/km sensitivity increases to about 5\%.
        The full DESI release expects to push sensitivity even lower than the current 2\%.
        The range of wavenumbers for which we have observations spans from $k\sim 0.001$ to about $k\sim 0.02$ s/km for eBOSS \citep{Chabanier+2019}, up to $k\sim0.025$ or $k\sim 0.04$ s/km (at $z=2.2$ and $z=3.8$ respectively) for early release DESI \citep{Ravoux+2023, Kara+2024}, and up to $k=0.2$ s/km for \citet{Karacayli:2022}.
        
        The shaded regions in Figure \ref{fig:PSratio_highz} represents the sensitivity derived from the eBOSS, DESI, and KODIAQ/SQUAD/XQ-100 surveys, assuming the \textit{SW Only} simulation represents ground truth. 
        The variations in the P1D do not appear large enough to be observable with current sensitivities.
        However, if DESI is able to push sensitivities down to 1.5\% variations due to the inclusion of jets may be visible. 
        The normalization shift due to the heating of \lya~absorbers is relatively small with the main changes to the P1D seen at $k>0.04$ s/km at $z<2.4$.
        These changes are due to the different $T_0$ evolution seen in the \textit{+Rad+Jet} simulation with a slightly warmer IGM.
        At $z=3.0$, $T_0$ is about the same for all four simulations, but by $z=2$ the IGM temperature in the no X-ray simulation is about $10\%$ higher, as shown in Figure~\ref{fig:thermalhistory}.
        This explains the changes in the P1D ratio of the \textit{+Rad+Jet} simulation, as the P1D is sensitive to the thermal history. It is important to note that cosmological parameter estimation marginalizes out the effect of the thermal history, so that this shift would likely not affect the derived cosmological parameters, even at the measurement accuracy of the full DESI survey. However, it implies that potentially detectable shifts in the thermal history at $z\sim 2$ could distinguish between some AGN feedback models. 
    
\section{Discussion}\label{s:Discussion}

    \subsection{Mean Flux Re-Scaling}\label{ss:meanflux}

        \begin{figure*}
            \centering
            \includegraphics[width = \linewidth, trim=0.0cm 0.0cm 0.0cm 0.0cm, clip=true]{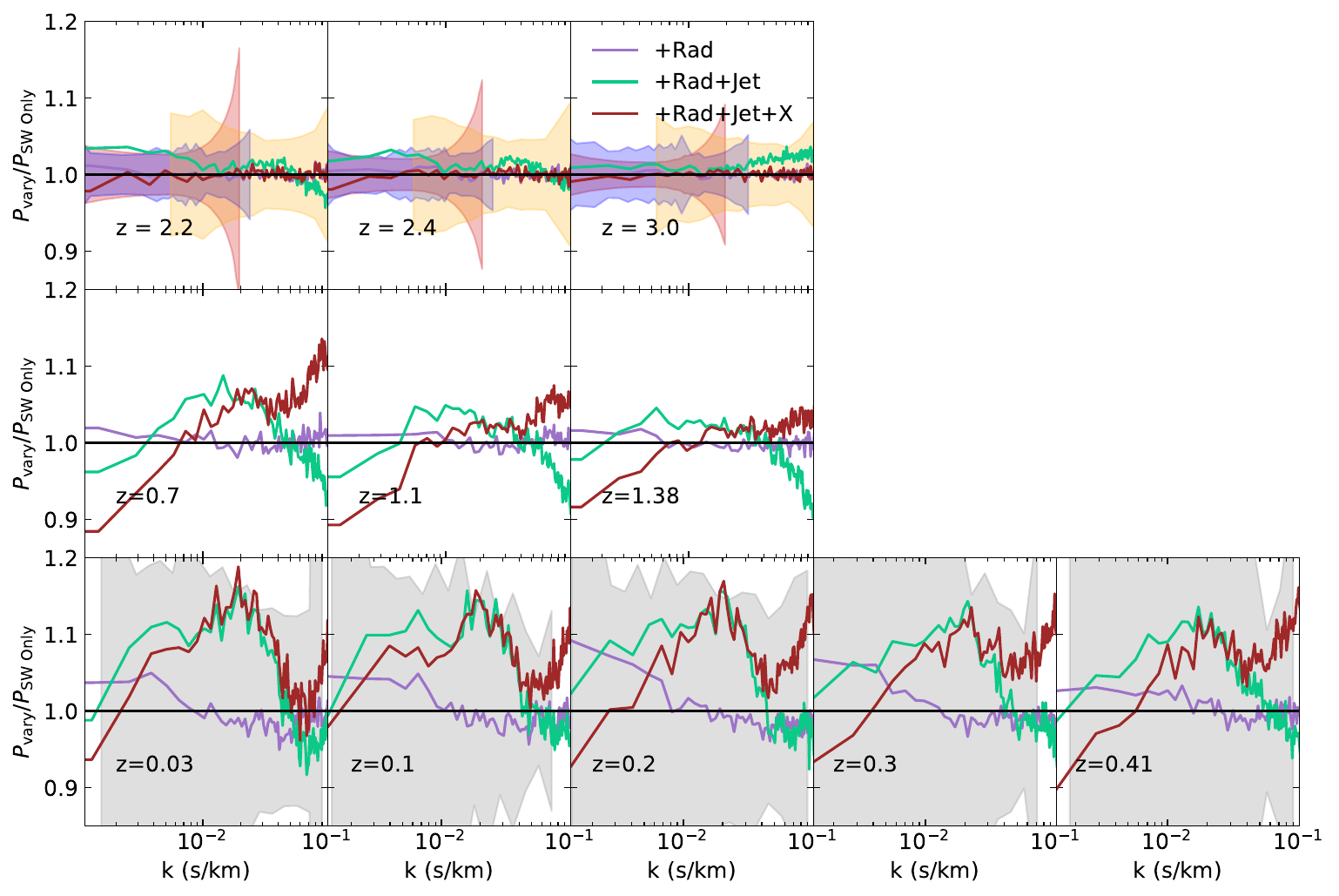}
            \caption{The top, middle, and bottom rows are the same as Figures \ref{fig:PSratio_highz}, \ref{fig:PSratio_interz}, and \ref{fig:PS_lowz} respectively but with the P1D re-scaled to have the same mean flux as the fiducial \simba simulation (\textit{+Rad+Jet+X}).}
            \label{fig:PSmeanflux}
        \end{figure*}

        For our main results we do not re-scale the effective optical depth ($\tau_\textnormal{eff}$).
        This is often done to account for uncertainties in the UVB model chosen but does not account for varying thermal histories.
        Since the only difference between the \simba simulations analyzed herein is the AGN feedback model implemented, we refrained from re-scaling $\tau_\textnormal{eff}$ to isolate the full physical effects of the AGN feedback.
        Re-scaling the P1D to the same mean flux mixes the different scales and makes it difficult to understand exactly where changes are coming from.
        However, this remains an important analysis for comparison with other works and to address degeneracies in the mechanisms that set the thermal state of the IGM. 

        Figure \ref{fig:PSmeanflux} shows all the redshift bins analyzed previously but with $\tau_\textnormal{eff}$ re-scaled to match the fiducial \simba simulation.
        When re-scaling $\tau_\textnormal{eff}$, changes to the shape of the P1D are largely preserved since the re-scaling acts to normalize the power.
        If the changes due to different AGN feedback models remain larger than observational precision after $\tau_\textnormal{eff}$ re-scaling, then the \lya~forest may prove a useful tool for constraining feedback models.
        
        For the low redshift bins, re-scaling the P1D to the same $\tau_\textnormal{eff}$ results in the simulations matching one another within current observational precision.
        However, some of the feedback variations change the P1D by $> 10\%$ and so the effects of feedback are potentially detectable in higher precision data.
        The intermediate redshift bins, $z=0.7 - 1.4$, change by  $\sim 10$\%.
        For higher redshifts the results are still within $\sim 4$\% but now the largest differences seen are at $k<10^{-2}$ s/km. 
        Notably, the differences at those $k$ values for redshift bins $z=2.2$ and $2.4$ are slightly larger than current observational precision, suggesting that the thermal history shift created by these AGN feedback models may be detectable with the full DESI survey.

    \subsection{The Role of AGN and Stellar Feedback}
    
        As previously stated, the most dramatic effects of AGN feedback on the \lya~forest P1D are seen at lower redshifts ($z\lesssim 0.7$). 
        The AGN jet feedback mode has the largest scale impact and it does not turn on until both the galaxy and SMBH have grown substantially. 
        The galaxy needs to have a certain stellar mass before it seeds a SMBH, and that SMBH needs to grow to a mass of $10^{7.5} M_\odot$ before it can even activate the jet mode.
        Therefore, it will take time before enough galaxies host SMBHs producing jet feedback to even be able to see those effects clearly in a summary statistic such as the P1D.
        
        We have mostly focused on the impact of AGN feedback, but as noted earlier, stellar feedback can play a varying role as more or less occurs in massive halos due to changes in AGN prescriptions.
        \citet{Tillman:2023AJ} saw that increasing the strength of the stellar feedback had a large impact on not only BH seeding in the \simba model (we still lack the observational constraints to improve this mechanism) but also on the growth of the SMBHs.
        This occurs when including the X-ray feedback mode with the change we see in the P1D, likely due to the complex interplay of SMBH growth and stellar feedback.
        However, it is interesting that we can see, in these simulations, the impact of stellar feedback heating directly when including the radiative feedback mode.
        In contrast, \citet{Tillman:2022} found essentially negligible effects on the column density distribution function from the radiative feedback mode.
        However, we note that the change we see in the P1D is at most 10\%, which would very likely be indistinguishable due to cosmic variance and observational uncertainty.

\subsection{Implications for observations.}

    At lower redshifts ($z<1$), small-scale structure signatures in the Lyman-$\alpha$ forest P1D is erased by AGN heating in the \simba simulation.
    This effect, rather than just reducing all P1D scales by a similar amount, is potentially differentiable observationally by non-Gaussian measures \citep[e.g.][]{Tohfa:2024}.
    If this is the case, it could provide an opportunity to disentangle the effects of AGN feedback from other degenerate effects such as the assumed ionizing background.
    More targeted studies of the \lya~forest exploring the space surrounding the most massive halos, which we might expect to host AGN jets, may also prove useful in isolating AGN feedback effects \citep[e.g.][]{Khaire:2024}.

    At high redshift ($z>2$), heating due to stellar feedback and AGN feedback likely cannot be distinguished from helium reionization heating using the P1D.
    However with the full DESI release predicted to increase observational precision, this may change.
    At these redshifts, it seems the temperature and ionization state of the \lya~forest is dominated by the assumed ionizing background via photoionization and photoheating. 
    However, this is in part by design of the AGN feedback model implemented in the \simba simulations.
    If through new observations we find AGN to be more prevalent at higher redshifts, then AGN feedback may play a larger role in setting the \lya~forest than seen here.
    Once observations are able to push to higher wavenumbers and improve their precision, revisiting the effect of AGN feedback on the P1D will be necessary. 
    Additionally, this analysis should be conducted for more AGN feedback models as the one implemented in these simulations may not be correct.
    
    Due to the small box size, we make no attempt to calculate any cosmological parameters to examine biases.
    Another study doing a similar analysis, and calculating those values, found that AGN feedback could bias the cosmological parameters inferred from \lya~statistics \citet{Chabanier:2020}.
    This work and \citet{Chabanier:2020} collectively looked at two different AGN feedback models.
    \citet{Chabanier:2020} found a 1-8\% effect on the low-$k$ end of the P1D from $z=4.25$ to $z=2$ respectively due to AGN, while in this work we find a $\lesssim 2\%$ effect on the P1D at $k<5\times 10^{-2}$ s/km and an 8\% effect at $k>5\times 10^{-2}$ s/km (although the largest effect is degenerate with a shift in the IGM mean flux).
    This implies that the \simba AGN radiative feedback model, the mode dominating at these redshifts, does not heat larger scales as much as HorizonAGN's quasar mode.
    There are a few differences between the HorizonAGN and \simba high accretion feedback models that may explain this. 
    In HorizonAGN, energy is stored when the SMBH accretes mass until enough energy is available to heat the gas surrounding the SMBH to $T\sim10^7$ K after which the energy is released thermally in a bubble surrounding the SMBH \citep{Dubois:2012}. 
    In \simba the high accretion mode is kinetic with a bipolar kick of gas within the BH kernal as high mass loading outflows \citep{SIMBA}. 
    The fact that HorizonAGN employs a thermal based quasar mode may explain why \citet{Chabanier:2020} saw more removal of power on large scales due to heating at higher redshifts. 
    In \simba, the kinetic based radiative mode does not produce heating on such large scales.
    For \simba, it is the jet mode that ejects heated gas but those jets will not be the dominant feedback mechanism until $z<2$. 
    Both the HorizonAGN and \simba quasar feedback modes efficiently quench galaxies, but the fact that they produce such different large-scale effects further emphasizes the necessity to better understand AGN feedback.
    
    Due to the differing conclusions between our studies and the uncertainty in AGN modeling, it is difficult to quantify just how much AGN feedback may affect the P1D at redshifts of $z>2$.
    However, the different predictions for AGN feedback effects on the P1D between simulations that utilize different AGN feedback models further emphasizes the potential of the \lya~forest as a constraint on feedback.
    It is clear from both studies that AGN feedback has at least a 1-2\% effect on the predicted P1D at low-$k$, although this is largely due to a shift in the mean IGM temperature in some models.
    If the full DESI release pushes precision to these levels, AGN feedback effects will become significant.
    Another factor to consider is SMBH seeding in simulations, an aspect that is motivated partly by observations but remains poorly constrained. 
    Observations from JWST at high redshifts will likely influence future SMBH seeding and growth prescriptions so the impact of AGN feedback on the \lya~forest could be affected.

    The impact of AGN feedback is reduced as redshift increases from 2, due to the lower number density of SMBHs.
    This implies the effects of AGN feedback are unlikely to impact constraints on the free-streaming length and mass of warm dark matter particles using the \lya~forest P1D at $z>4$.
    Higher redshifts are favored for these studies in part due to HeII reheating being unlikely to affect the thermal history until $z<4$ \citep{Villasenor:2023}.
    However, it is possible that other AGN feedback models with more impactful heating at high-$z$, such as the model in HorizonAGN, may still alter the constraints from these studies.
    Alternatively, the use of higher resolution quasar spectra at $z>5$ may prove useful in determining the extent of AGN feedback effects \citep[see, e.g.~][]{Gaikwad:2020}.

\section{Conclusion}\label{s:Conclusion}

    With the goal of better understanding how AGN feedback may impact the observed \lya~forest across a wide range of redshifts ($0.03<z<3$), we study the P1D for different \simba simulations that vary the AGN feedback model implemented.
    Comparing these variations to the stellar feedback only (\textit{SW Only}) \simba simulation allowed us to analyze the effects of individual AGN feedback modes on the \lya~forest as they are included.
    Our main findings are:
    \begin{itemize}
        \item The AGN jet feedback mode has the most dramatic effect on the \lya~P1D,  consistent with previous work \citep{Christiansen:2020, Tillman:2023AJL, Tillman:2023AJ}.
        \item AGN feedback effects on the P1D can operate on all scales from pure ionization of hydrogen through heating (causing the erasure of structure on all scales), altering the thermal evolution of the IGM (pressure smoothing suppressing high-$k$), and thermal broadening of absorbers (suppressing high-$k$).
        \item The suppression of the high-$k$ end of the P1D due to thermal broadening is stronger as $z$ decreases and likely becomes observable around $z\sim 1$. 
        This implies that additional observational data in the redshift range of $0.5<z<1.8$ may help distinguish between different AGN feedback models and their effects on the \lya~forest. 
        \item We find the effects of AGN feedback on the P1D at higher redshifts ($z>2.0$) to be $\sim2$\% at $k<2\times 10^{-2}$ s/km and $\sim5-8$\% for higher $k$. 
        Given these results, the fiducial \simba model does not significantly alter the $z>2$ \lya~forest. 
        However the model that removes X-ray feedback (\textit{+Rad+Jet}) does alter the \lya~forest through a change in IGM temperature.
        \item Re-scaling the \lya~forest to the same mean transmitted flux reduces differences between predicted P1Ds to $<20\%$ at lower redshifts ($z<1$). 
        However even with this re-scaling, P1D observational data with precision as high as 10\% would be sufficient to distinguish between these simulations.
        \item Larger box simulations will be necessary to better understand the impact AGN may have on the \lya~forest. 
        These simulation must be large enough to more accurately sample the most massive halos, have sufficient resolution to study the IGM across a wide range of redshifts ($0<z<4$), and explore various AGN feedback implementations (see Appendix for more details).
    \end{itemize}

    With the available observational data from eBOSS \citep{Chabanier+2019} and the more recent DESI \citep{Ravoux+2023, Kara+2024} surveys, it is important to consider galaxy formation effects on these observations, especially in the era of precision cosmology. 
    We show that the default \simba AGN feedback model has a small effect on the \lya~forest P1D at $z>2$, so that \simba-like simulations are consistent with observational data. 
    One of the feedback variations considered changes the P1D via the temperature-density relation of the IGM. We show explicitly that the effect on small scales is reduced if both simulations are rescaled to the same observed \lya mean flux. 
    Our results differ from those of \cite{Chabanier+2019} and \cite{Viel:2013}, because the \simba AGN feedback model is less aggressive at $z>2$ than considered in those works. 
    In particular, the OWLS-AGN simulation from \cite{Viel:2013} under-predicts the galaxy stellar mass function by an order of magnitude.
    Since the effects of AGN feedback on cosmological studies using the \lya~forest P1D varies between feedback models, further constraining which AGN feedback models are most realistic will be necessary to properly account for such baryonic effects. 
    This will require simulation boxes big enough to obtain sufficient halo sampling at all masses but still with a high enough resolution to study the IGM.

    At lower redshifts the \simba simulation AGN feedback model has a much more dramatic effect on the \lya~forest P1D.
    These effects become observable within 25\% accuracy at $z\sim 1$.
    However, rescaling the \lya~forest P1D to the same mean transmitted flux reduces the difference between P1D to $<20$\%.
    This implies that the effects of AGN feedback are still highly degenerate with the assumed UVB for the \lya~forest P1D.
    Given current observational data, it is difficult to differentiate the \simba AGN feedback model from other models using the P1D.
    However if we can improve observational precision another $\sim10$\% it may become possible to use the \lya~forest P1D as a constraint for the \simba AGN feedback model at $z<0.5$.
    Alternatively, the effects of feedback within the intermediate redshifts that we do not currently have observational data for ($0.5<z<1.8$) is a promising direction of study.
    Especially given the opportunity of a new space-based FUV spectrograph with Habitable Worlds Observatory\footnote{\url{https://science.nasa.gov/astrophysics/programs/habitable-worlds-observatory/}} , predictions for observations in that redshift range will be valuable.

\begin{acknowledgments}
M.T.T. thanks Daniele Sorini for helpful discussions about the role of AGN feedback in the \simba simulations in regulating star formation. M.T.T. thanks Romeel Dave for insights on the intended purpose and motivation for the different AGN feedback modes in \simba. M.T.T. thanks Doug Rennehan for useful discussions on the \simba AGN feedback model and particle resolution. 
B.B. acknowledges support from NSF grant AST-2009679.
B.B. and S.B. acknowledge support from NASA grant No. 80NSSC20K0500. This research was also supported in part by the National Science Foundation under Grant No. NSF PHY-1748958.
B.B. is grateful for generous support from the David and Lucile Packard Foundation and the Alfred P. Sloan Foundation.
G.L.B. acknowledges support from the NSF (AST-2108470 and AST-2307419, ACCESS), a NASA TCAN award, and the Simons Foundation through the Learning the Universe Collaboration.
The authors thank the Flatiron Institute Center for Computation Astrophysics for providing the computing resources used to conduct much of this work. The Flatiron Institute is a division of the Simons Foundation.
\end{acknowledgments}

\appendix

\section{Box Size - Cosmic Variance and Particle Resolution}\label{ss:convergence}
     
    All of the simulations analyzed herein utilize the same random seed, and the 50 Mpc/$h$ boxes have a P1D converged within $\sim5-10$\% of the flagship 100 Mpc/$h$ \simba simulation.
    Results from \citet{Bird:2023} found cosmic variance due to box size (specifically between 120 Mpc/h to 60 Mpc/h box lengths) for $z<3.6$ is at the level of 2\%, but for the highest modes ($\sim 50$ Mpc/h scales) this can reach up to 5\%.
    \simba is less converged than the simulations presented in \citet{Bird:2023}, as it uses a box size a factor of $2.4$ smaller, but it seems more likely that a change in sub-grid physics such as feedback is the culprit. 
    The AGN jets can travel distances on the scales of Mpc and are typically hosted in larger halos, so a large box with a well sampled halo mass function is likely required to ensure convergence of the \simba AGN sub-grid model.
    The largest difference seen between the two \simba boxes are seen at lower redshifts, where AGN feedback dominates, and at low-$k$, where we expect differences due to box size.
    The effects on the P1D seen here, which are $\leq10$\% for $z<2.0$ and $\leq5$\% for $z>2.0$ may be upper bounds when simulated in larger halos.
    At $z<1.0$, the effects of the feedback prescription are greater than 10\%, but at higher redshifts, $z > 2$, the largest changes seen in the P1D are closer to 2\% and may be reduced in a larger box.
    Since the simulations being studied use the same initial conditions, and change only the AGN feedback implementation, it is still possible to analyze trends caused by feedback even if the simulations are not completely converged with box size. 
    It is clear that AGN feedback has an effect since the simulations isolate and vary only the AGN model, but the magnitude of this effect could change based on the SMBHs sampled by the implemented random seed.
    
    The box size of the full \simba run, $100$ Mpc/$h$, contains limited numbers of the most massive halos $M_h \gtrsim 10^{14} M_\odot$.
    This in turn limits the sample size of SMBHs and thus the impact of AGN feedback.
    In \simba, at lower redshifts ($z<2$) AGN feedback is efficient at evacuating baryons from halos of $M_h = 10^{12}$ to $10^{13} M_\odot$ and cosmic variance at the high-mass end ($M_h \geq 10^{13} M_\odot$) of the halo mass function shows a variance of $\sim 30\%$ at $z=2$ to $\sim15$\% by $z=0$ \citep{Sorini:2022}.
    For example, at $z=0.1$ a 15\% variance in the abundance of the most massive halos could plausibly translate to 15\% variance in the number of jet producing AGN, and so imply an effect of AGN jets on the predicted P1D of around $60\pm 9$\%.
    Whether or not this trend actually holds if halo masses are fully sampled is not clear.
    It is also possible those SMBHs do not produce enough energy to evacuate the most massive halos in a similar way as the $10^{13} M_\odot$ halos, which would therefore cause only minor changes to the P1D.
    Our results suggest that running bigger simulation boxes with full hydrodynamical astrophysics will be necessary to further constrain AGN feedback models.
    \lya~forest convergence studies have been conducted on simulation box size \citep[such as in][]{Bird:2023}, but it would be beneficial to expand on these with even bigger boxes such as with MilleniumTNG \citep{MilleniumTNG:2023}.

    Sufficient particle resolution is also necessary for accurately predicting the P1D since larger values of $k$ probe small physical scales.
    If fluctuations in flux on small physical scales are artificially suppressed due to lack of resolution, then fluctuations on larger scales will increase, meaning the overall P1D will be un-physically shifted to lower $k$ \citep{Lukic:2015}. 
    The ideal way to check for convergence with particle mass resolution is to compare to a higher resolution run of the same simulation.
    The $100$ Mpc/h and $50$ Mpc/h \simba simulations have the same particle resolution; however, there exists a ``high-res'' \simba run  with 2$\times 215^3$ particles  and a box with side length 25 Mpc/h.
    Comparing the P1D of the simulations herein to that of the high-res \simba simulation reveals a $\sim 40$\% difference in the P1D at $z = 0.1$.
    This difference is largest at lower $k$ values and redshifts, and the effects naturally diminish as redshift increases, behaving similar to that of the AGN feedback effects.
    By $z=3.0$ the difference between the simulations analyzed herein and the high-res simulations drops to $\sim 15$\%.
    Because the disparity between the runs is reduced at higher $k$ values, where resolution should have the most impact, and at higher $z$, where IGM resolution is more important, the difference is likely to be due to lack of convergence in the AGN model that heats the diffuse gas rather than a lack of convergence in modeling the hydrodynamics of the diffuse gas itself.
    
    In fact, this explanation can be checked by re-scaling the effective optical depth in each of these simulations to the same value.
    This is an approximate correction to help account for differences in the assumed UVB, specifically the HI photoionizing rate $\Gamma_{HI}$, between simulations.
    $\Gamma_{HI}$ does not change between these simulations but this correction will help account for differences in the neutral fraction of hydrogen.
    After the correction, the predicted P1D between simulations match within 5\% at high $k$ and within 10\% for the lowest modes (which is due to the much smaller box size of the high-res simulation).
    This implies a change in neutral hydrogen distribution most likely caused by the AGN feedback model behaving different in a higher resolution simulation, particularly the jets as they appear to have the greatest affect on the IGM.
    This could be due to an under-sampling of high mass SMBHs producing AGN jets in the smaller box, but with a change as dramatic as 40\% it is more likely to be due to a difference in the behavior of jet particles in a high resolution environment.
    This highlights a struggle in testing feedback sub-grid models at different box sizes and resolutions.
    If the feedback model was originally tuned for different conditions, it is difficult to disentangle what changes are due to sampling and those due to differences in the sub-grid models.
    
    Another way we can check if the resolution in the 50 Mpc/h $2\times512^3$ particle box is sufficient is based on IGM theory and approximate calculations.
    Resolving the filtering scale using multiple resolution elements is necessary to achieve convergence for the P1D.
    The exact filtering scale depends on the full thermal history of the simulation but an approximation can be made using the Jeans' length. 
    Post-reionization, the filtering scale at one redshift is equal to the Jeans' scale at an earlier time \citep{Gnedin:2000}. 
    In terms of the line of sight velocity, we can write the Jeans' scale for IGM gas traced by the \lya~forest with a mean temperature $T_0$ and an overdensity $\Delta$ as 

    \begin{equation}
        \sigma_J \approx 77.1 \;\textnormal{km s}^{-1} \sqrt{\frac{T_0}{10^{4}\textnormal{K}}} \Delta^{\gamma/2 -1}.
    \end{equation}
    
    \noindent The equation above uses the power law relation $T=T_0\Delta^{\gamma-1}$ that the \lya~forest is expected to follow and assumes a mean molecular weight of $\mu=0.61$ for a mixture of ionized hydrogen and slightly ionized helium \citep{Nasir:2016}. 
    The filtering scale will then go as $\sigma_p = f_J \sigma_J$ where $f_J < 1$ and is generally smaller closer to reionization \citep{Gnedin+Hui:1998}. 
    A reasonable expectation for $z\sim 3$ is for $f_J \approx 0.65-0.4$and $f_J$ increases with lower redshifts.
    Thus $f_J = 0.4$ is a reasonable lower bound at $z\sim3$ and certainly lower than necessary at lower redshifts.
    
    To calculate the filtering scale the gas is assumed to have a temperature of $T_0=10^4$ K (a reasonable approximation for $z<4$) and overdensity $\Delta=1$. 
    $T_0$ decreases with redshift after helium reionization and $\Delta$ for the \lya~forest is expected to be higher at lower redshifts. 
    The effects of these changes on the calculated filtering scale counteract each other to some degree (how much?).
    To check for convergence, the filtering scale will be compared to the average size of the numerical fluid particles expected to be contributing to the \lya~forest. 
    Since not all particles are the same size in the \simba simulations, we determine the average size of the IGM particles of interest by looking at the radii of particles with temperatures and densities within the range that the power law tail resides on the temperature-density plane (e.g.\ particles with $T\lesssim 10^{4.5}$ K and $n_{H} \lesssim 10^{-5}$ cm$^{-3}$ at $z=0.1$).
    
    Using these assumptions at $z=0.1$ the wavenumber at which pressure smoothing dominates at is $k_p = 0.049$ s/km, corresponding to a physical distance $L_p \approx 720$ kpc (based on the assumptions made for this calculation, this is actually expected to be larger).
    The average distance between IGM particles in the simulation is $\approx 180$ kpc.
    Since we are resolving multiple elements within the filtering scale using a conservative estimate, resolution should be sufficient at this redshift.
    At higher redshifts the filtering scale is smaller so resolution is more important. 
    By $z=1.3$ the number of elements resolved per filtering scale is down to $\sim 2.7$ and by $z=3.0$ this lowers to $\sim 1.2$.
    With these results it is clear that the resolution of the \lya~forest in the feedback variant simulations should be sufficient at lower redshifts $z\lesssim1.0$ but may suffer at higher $z$.
    
    As a result of this calculation we find that the simulations analyzed herein should have a sufficiently resolved IGM to trust the generated \lya~forest P1D, at least for the lowest redshift bins analyzed.
    This conclusion is consistent with the empirical test in \cite{Burkhart_2022} and our findings of convergences of 5-10\%.
    These results also further motivate that the dramatic lack of convergence between the high-res \simba simulation and the feedback variant simulations at low-$z$ is unlikely to be due to the resolution of the IGM but rather a difference in the behavior of sub-grid models between resolutions.
    Indeed the lack of convergence is worse at lower $z$, where AGN feedback is most impactful, and resolves as $z$ increases to $2-3$, where AGN impacts on the P1D is minimized.
    At least in the IGM, the effects of the AGN jet feedback sub-grid model appear somewhat dependant on particle resolution.
    
\bibliography{mybib}{}
\bibliographystyle{aasjournal}

\end{document}